\documentclass[slac_one]{revtex4}
\usepackage{graphicx}
\usepackage{fancyhdr}
\newcommand{\met}{/\!\!\!\!E_{T}}
\begin{document}

\title{{\small{Hadron Collider Physics Symposium (HCP2008),
Galena, Illinois, USA}}\\ 
\vspace{12pt}
Search for Technicolor Particles\\ Produced in Association with W Boson at CDF} 

%

\author{Y. Nagai, T. Masubuchi, S. Kim}
\affiliation{University of Tsukuba, Tsukuba, Ibaraki, Japan}
\author{W.-M. Yao}
\affiliation{LBNL, Berkeley, CA, USA}

\author{on behalf of the CDF Collaboration}

\begin{abstract}
We present a search for technicolor particles decaying into $b\bar b$, $b\bar c$ or $b\bar u$
and produced in association with $W$ bosons in $p\bar p$ collisions at
$\sqrt{s}= 1.96\,\mathrm{TeV}$. The search uses approximately
$1.9\,\mathrm{fb}^{-1}$ of the dataset accumulated in the CDF II detector
at the Fermilab Tevatron.
We select events matching the $W$ + 2-jets signature and require at least one jets to be identified as $b$-quark jets. 
In the case of exactly one vertex $b$-tagged events, we apply a neural network
flavor separator to reject contamination from charm and light quark jets. 
The number of tagged events and the invariant mass distributions of $W+2$
jets and dijets are consistent with the Standard Model expectations.
We succeed to set a large $95\%$ confidence level excluded region 
on the $\pi_{T}$ mass v.s. $\rho_{T}$ mass plane.
\end{abstract}

\maketitle

\thispagestyle{fancy}


\section{INTRODUCTION} 
\begin{figure*}[t]
\centering
\includegraphics[width=0.4\textwidth]{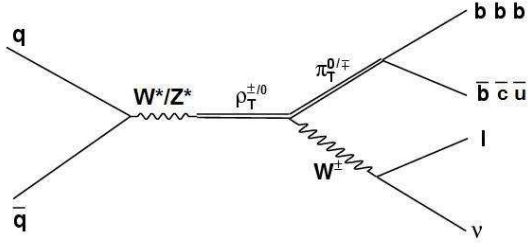}
\caption{Feynman diagram for $p\bar p \rightarrow \rho_T^{\pm/0}
\rightarrow W^\pm \pi_T^{0/\mp} \rightarrow \ell \nu b\bar b$, 
$\ell \nu b\bar c$ or $\ell \nu b\bar c$ production. } \label{Feynman}
\end{figure*}

\begin{figure}[htbp]
  \begin{center}
    \includegraphics[width=0.3\textwidth]{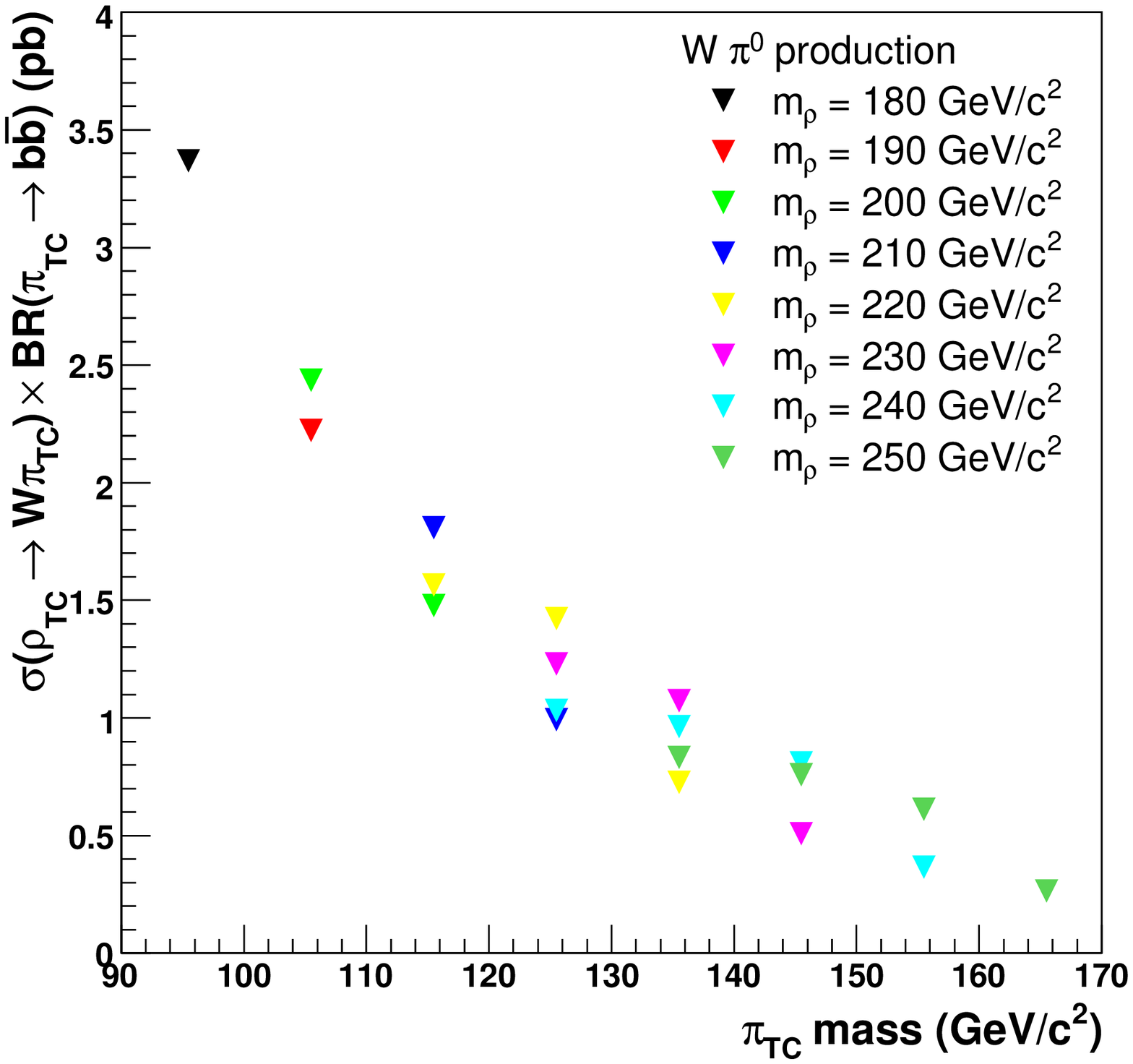}
    \includegraphics[width=0.3\textwidth]{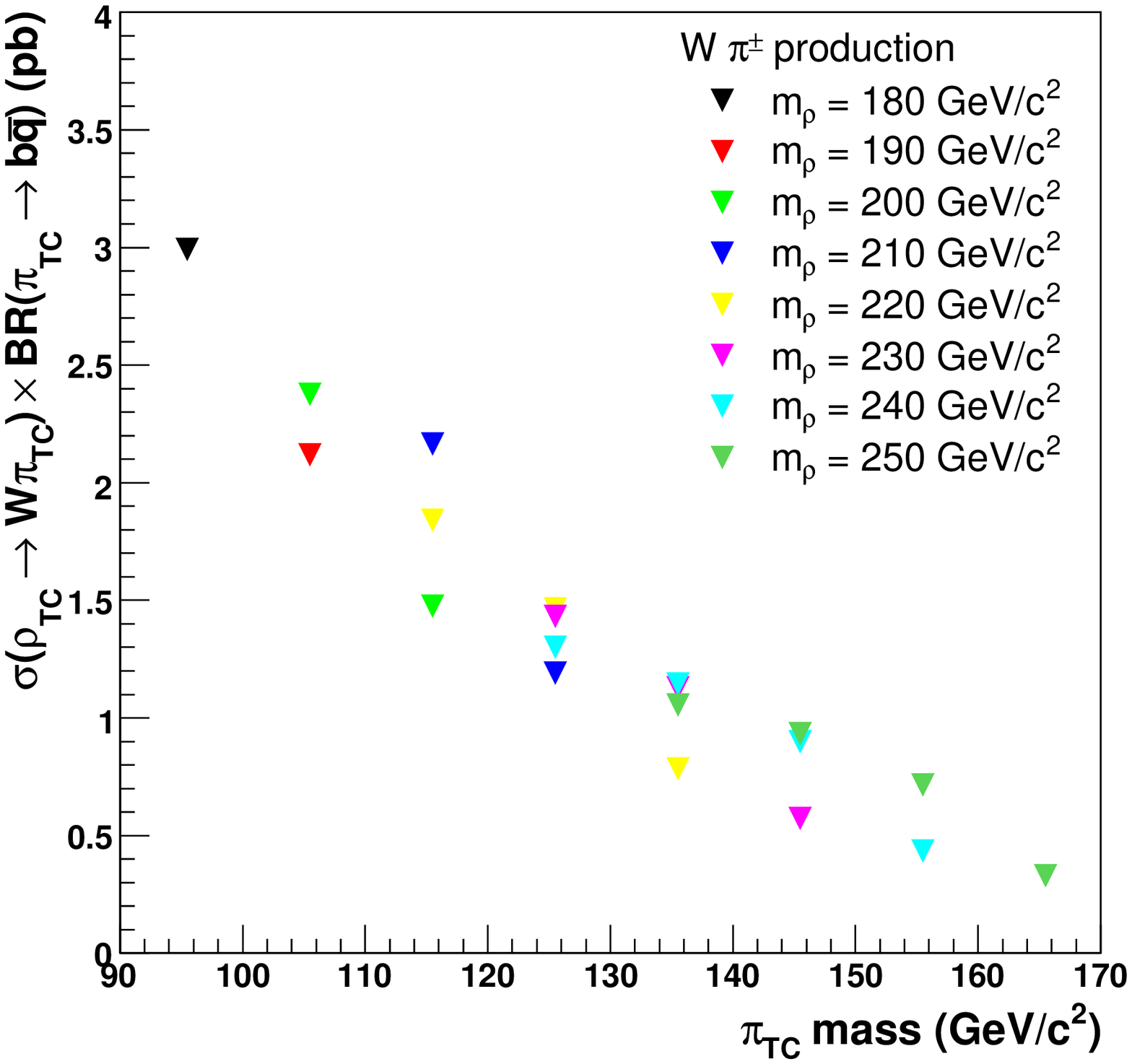}
    \caption{Production cross section calculated in {\sc Pythia} 6.216 with the Technicolor Straw Model (TCSM).
      (Left) $p\bar p \rightarrow \rho_T^\pm \rightarrow W^\pm \pi_T^0 \rightarrow \ell \nu b\bar
      b$ production cross section as a function of the $\pi_T$ mass for various $\rho_T$ masses.
      (Right) $p\bar p \rightarrow \rho_T^0 \rightarrow W^\pm \pi_T^\mp \rightarrow \ell \nu b\bar c$, $\ell \nu b\bar u$
      production cross section as a function of the $\pi_T$ mass for various $\rho_T$ masses.}
     \label{fig:cross_section}
  \end{center}
\end{figure}

The mechanism of electroweak symmetry breaking in the standard model
is still unknown. 
The most popular mechanism to induce
electroweak symmetry breaking of the gauge theory, resulting in
massive gauge bosons and fermions, are the Higgs mechanism~\cite{Higgs:1964pj}.
Alternatively, there is a theory which induces the electroweak symmetry breaking by dynamically.
This is a technicolor theory~\cite{lane} which predicts the existence of the new strong force
(technicolor) and new fermions (technifermions), and does not require elementary scalar bosons.
Technicolor interact between technifermions to form the bound states (technihadrons)
such as $\rho_{T}^{0,\pm}$, $\pi_{T}^{0,\pm}$ and $\omega_{T}^{0}$, 
analogous to the mesons on QCD.
In $p\bar p$ collisions at Tevatron, one of the most likely process is 
$\rho_{T} \rightarrow W\pi_{T} \rightarrow l^{\pm} \nu b\bar{b}, b\bar{c}$ or $b\bar{u}$,
depending on their charge (Figure~\ref{Feynman}).
In this analysis, we focus on this process with the Technicolor Straw Man Model (TCSM)~\cite{tcsm}. 
With assuming the TCSM, the production cross secion of these processes is the order picobarns
(Figure~\ref{fig:cross_section}).  

\section{DATA SAMPLE \& EVENT SELECTION}
For this analysis, we use a data sample corresponding to approximately 
$1.9\,\mathrm{fb}^{-1}$ of integrated luminosity accumulated in the CDF II
detector at the Fermilab Tevatron~\cite{CDFdetector}.
The events are collected with high-$p_T$ electron or muon triggers,
which can detect electrons or muons with $E_T$ or $p_T > 18 \mathrm{GeV}$. 
We further require the electron or muon that is isolated with $E_T$ or $p_T > 20 \mathrm{GeV}$ at offline level.

To select the $W$ + 2jets final state, we require that events have the large missing transverse energy and one or two $b$-jets. 
Therefore, we require that events have $\met > 20 \mathrm{GeV}$ and exactly two jets,
where jets are defined using a cone algorithm with radius $0.4$. 
We count jets which have $E_T > 20 \mathrm{GeV}$ and $|\eta| < 2.0$.

To reduce the background, we require that at least one jet in the event are identified as $b$-jet 
by the Secondary Vertex Tagging Algorithm.
The secondary vertex tagging algorithm identifies $b$-jets by fitting tracks 
displaced from the primary vertex.
This method has been used by other analyses at CDF~\cite{Top,Higgs}.
In Addition, we add the Jet Probability Tagging Algorithm that identifies $b$-jets by requireing 
a low probability that all tracks contained in a jet originated from the primary vertex, 
based on the track impact parameters~\cite{jetprob}. 
To be considered for double tag category, we require that events have two secondary vertex tagged jets 
or one secondary vertex tagged jet and one jet probability tagged jet.

To increase the signal acceptance, we also make use of the exactly one $b$-tagged events 
with secondary vertex tagging algorithm. 
For the exactly one $b$-tagged events, 
we apply the neural network $b$-tagging algorithm to improve signal-to-background ratio 
by separating $b$-jets from $c$-jets or light flavor jets. 
This neural network $b$-tagging algorithm is used in previous analysis~\cite{Higgs}. 
With using this algorithm, we can improve the purity of $b$-jets while keeping about 90\% signal.

The dijet mass is reconstructed from the 2 jets in selected events.
To reconstruct the $W$ + 2jets invariant mass, we need to determine the
$p_z$ of the neutrino from the $W$ boson. After using the $W$ mass
constraint to solve for the kinematics of the $\ell\nu$ system, we
take a smaller value of the two $p_z$ solutions. (If there is no real
solution for $p_z$, we take the real part of the complex solution.)

\section{TECHNICOLOR SIGNAL SAMPLES}
The signal samples are generated with the {\sc Pythia} program~\cite{pythia}.
{\sc Pythia} version 6.216 implements the TCSM~\cite{tcsm} 
in leading-order calculations.
We set the mass parameters as $M_V = M_A = 200\,\mathrm{GeV}/c^2$, 
the charge of up-type technifermion as $Q_U = 1$ and the mixing angle between isotriplet technipion 
interaction and mass eigenstates as $sin \chi = 1/3$ in this model. For the other parameters on TCSM, 
we use the default value in {\sc Pythia}.
For this study, we focus on the mass region as: $m(W) + m(\pi_{T}) < m(\rho_{T}) < 2 \times m(\pi_{T})$,
$180 \mathrm{GeV}/c^2 < m(\rho_{T}) < 250 \mathrm{GeV}/c^2$ 
and $95 \mathrm{GeV}/c^2 < m(\pi_{T}) < 165 \mathrm{GeV}/c^2$
because of the kinematical threshold of $W \pi_T$ production and pair $\pi_T$ production.

For the systematic uncertainties of signal acceptance, 
we consider the effects from lepton identification, trigger, the $b$-tagging efficiency, initial and
final state radiation effects (ISR/FSR), parton distribution function (PDF) and the jet energy scale (JES).
Lepton identification uncertainty is less than 2\%, trigger uncertainty is less than 1\%, 
ISR/FSR uncertainty is 1.8-11.6\%, JES uncertainty is 2.7-11.3\%, PDF uncertainty is 2.3-3.8\% 
and b-tagging uncertainty is 4.3-17.0\%, respectively.

\section{BACKGROUNDS}
This analysis builds on the method of background estimation detailed
in Ref.~\cite{Top}.  In particular, the contributions from the
following individual backgrounds are calculated: falsely $b$-tagged
events (mistag), $W$ production with heavy flavor quarks ($W+bb$, $W+cc$, $W+c$), QCD events with
false $W$ signatures (NonW), top quark pair or single production ($t \bar{t}$, Single Top)
and diboson production ($WW$, $WZ$, $ZZ$).

We estimate the mistag events by using the mistag probability that measured from the inclusive jets sample.
Such mistag rate are obtained using negative tags, 
which are the tags that appear to travel back toward the primary vertex.
The mistag rate derived from negative tags is due to tracking resolution
limitations, but they provide a reasonable estimate of the number of
false positive tags after a correction for material interactions and
long-lived light flavor particles.

The number of events from $W$ + heavy flavor is calculated using
information from both data and Monte Carlo samples. We calculate the
fraction of $W$ events with associated heavy flavor production in the
ALPGEN Monte Carlo program interfaced with the PYTHIA parton shower
code~\cite{pythia,alpgen}. This fraction and the tagging efficiency
for such events are applied to the number of events in the original
$W$+jets sample after correcting for the $t\bar t$ and electroweak
contributions.

We constrain the number of QCD events with false $W$ signatures by
assuming the lepton isolation is independent of $\met$ and measuring the
ratio of isolated to non-isolated leptons in a $\met$ sideband region.
The result in the tagged sample can be calculated in two ways: by
applying the method directly to the tagged sample, or by estimating
the number of non-W QCD events in the pretag sample and applying an
average $b$-tagging rate.

\section{RESULTS}
We perform a direct search for a resonant mass peak
in the reconstructed $W+2$-jets and dijet invariant mass distributions 
from the single-tagged and double-tagged events.

\begin{figure}[htbp]
  \begin{center}
    \includegraphics[width=0.25\textwidth]{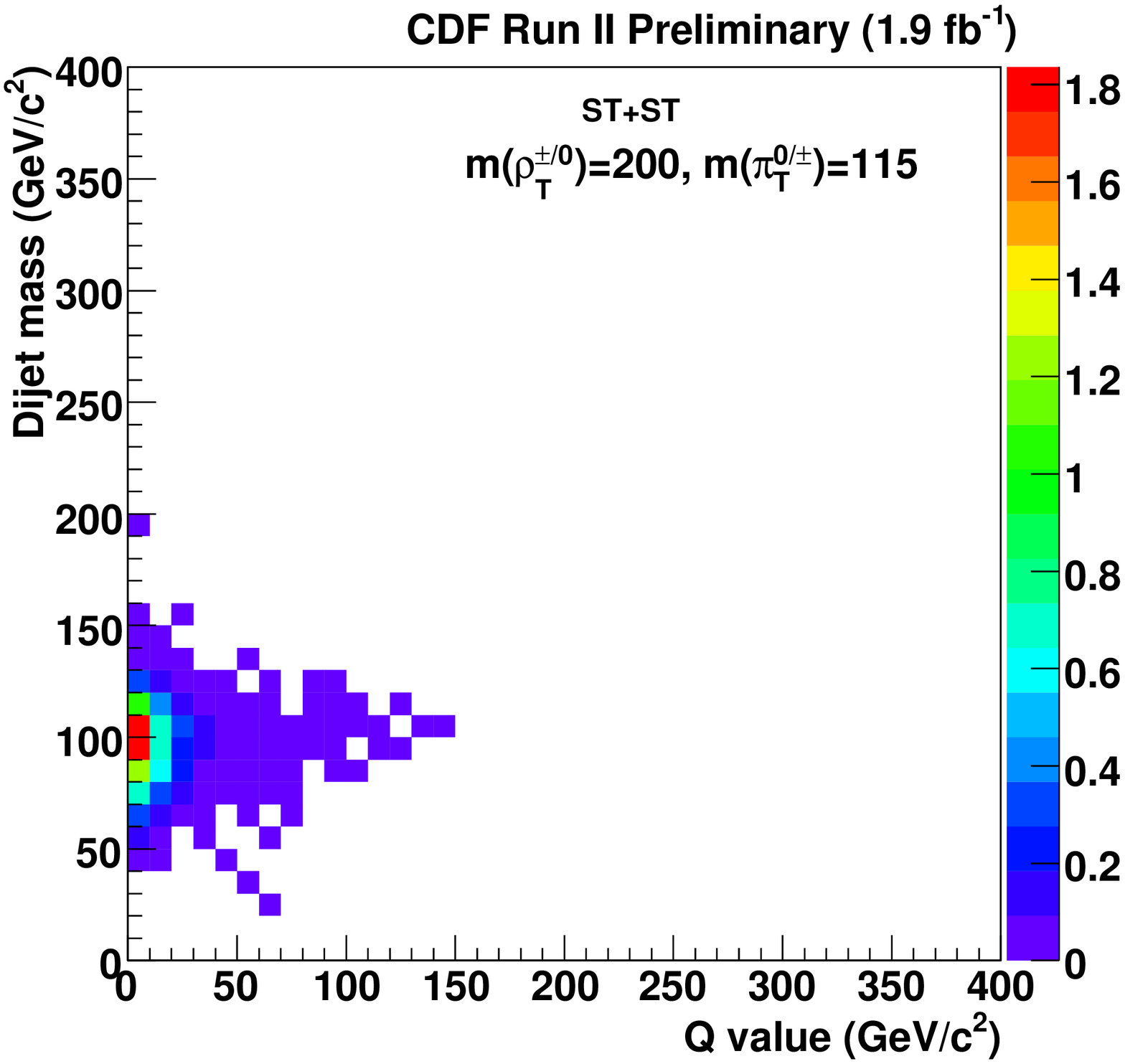}
    \includegraphics[width=0.25\textwidth]{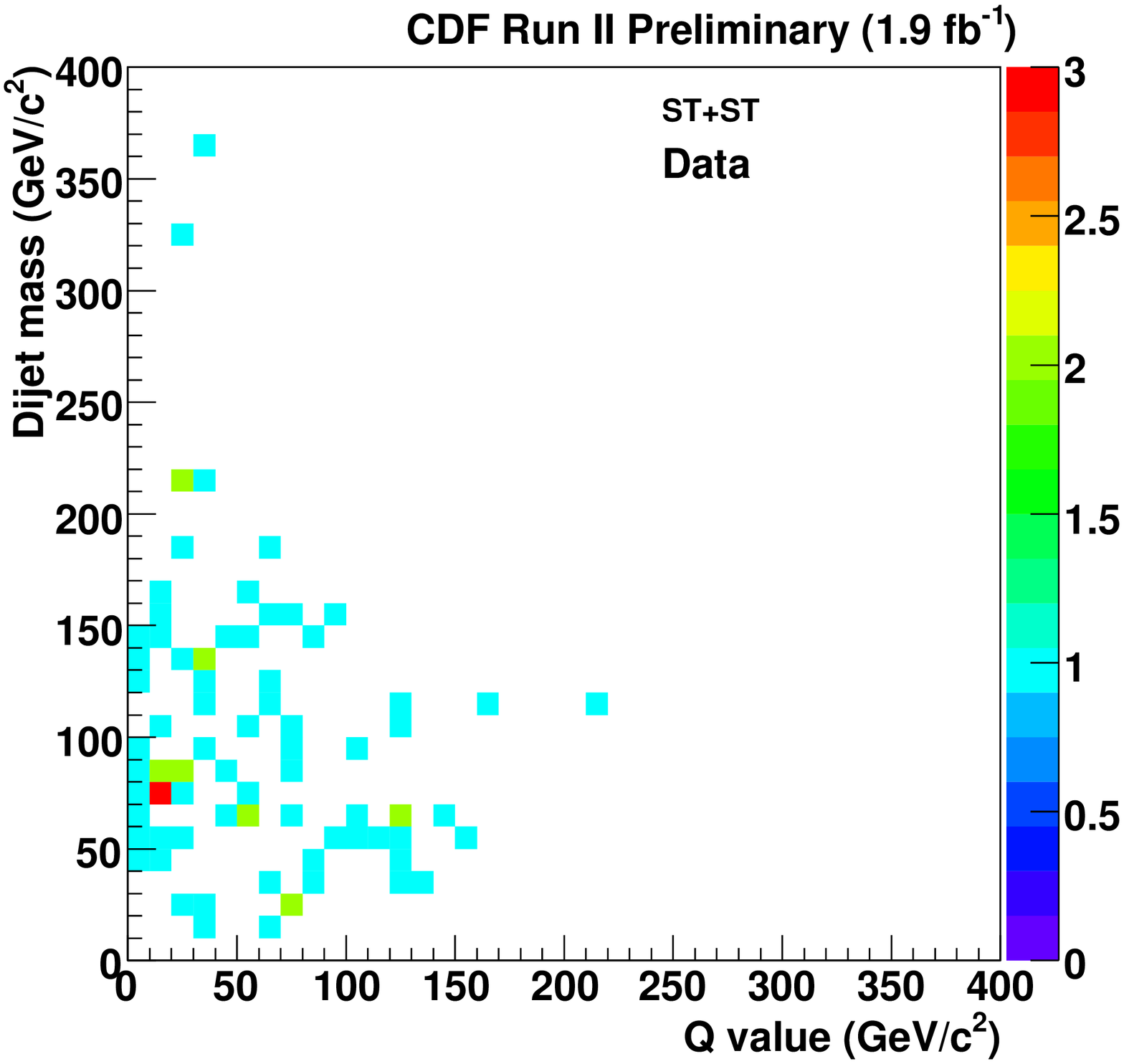}
    \includegraphics[width=0.25\textwidth]{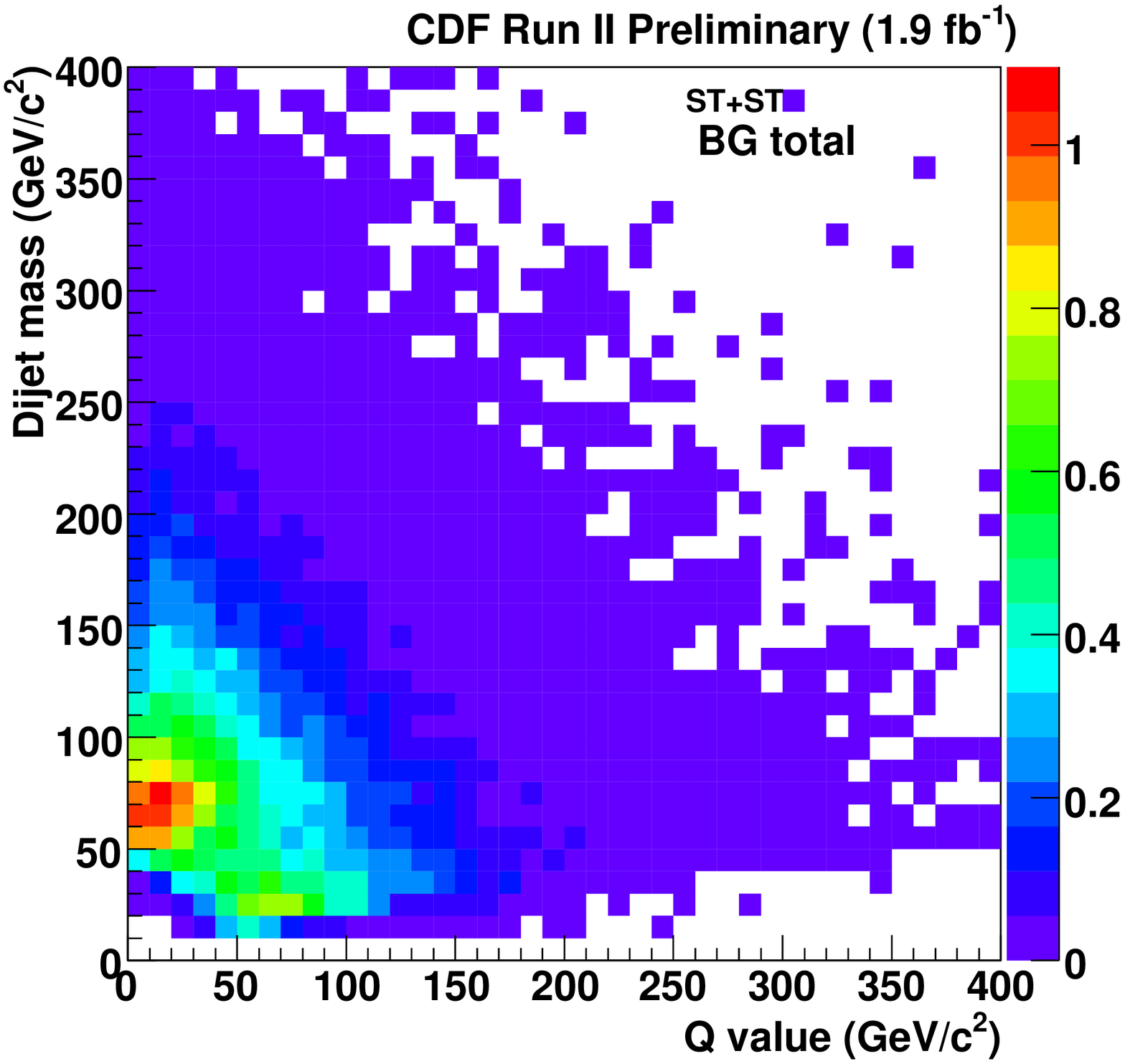}
    \includegraphics[width=0.25\textwidth]{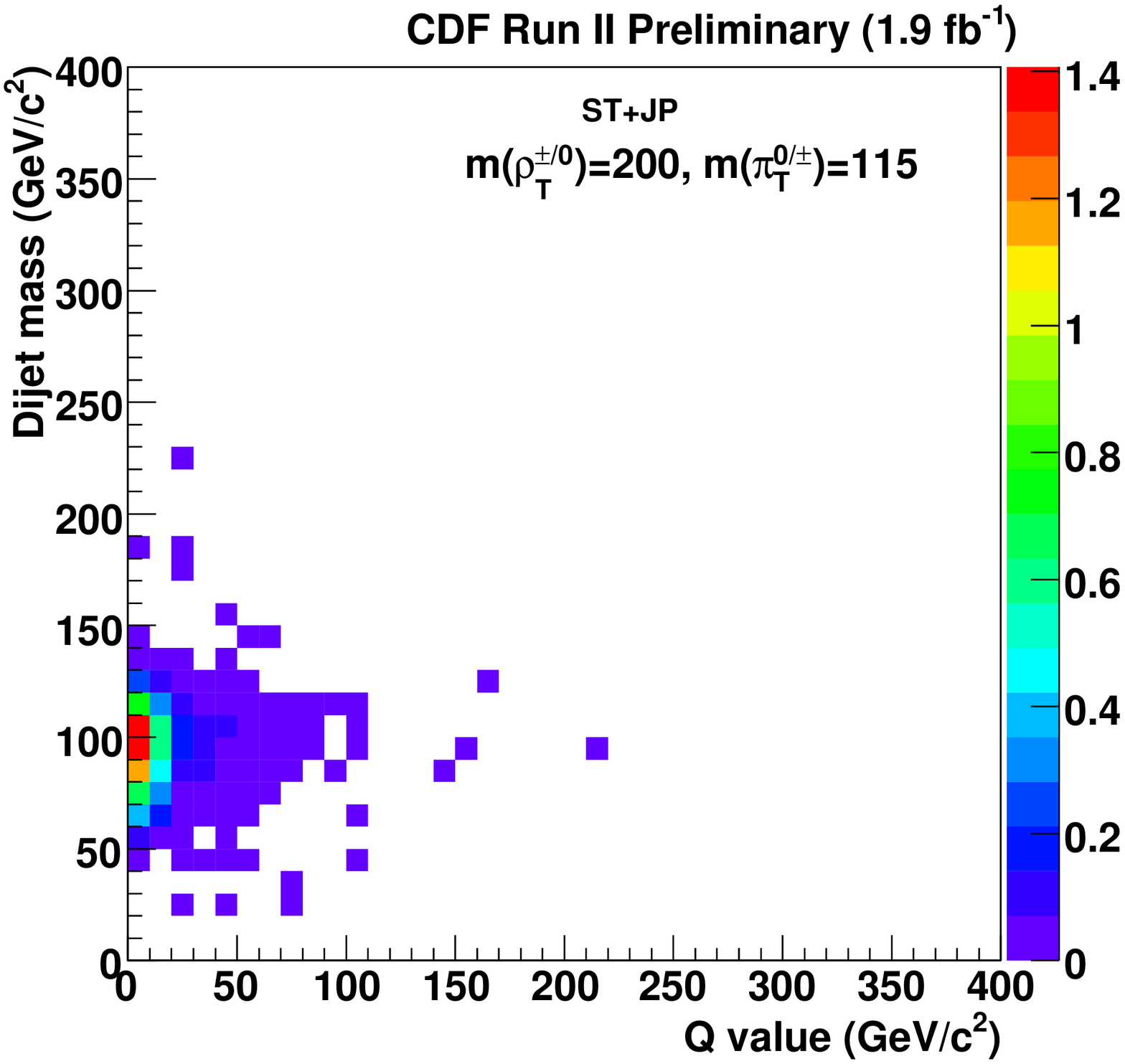}
    \includegraphics[width=0.25\textwidth]{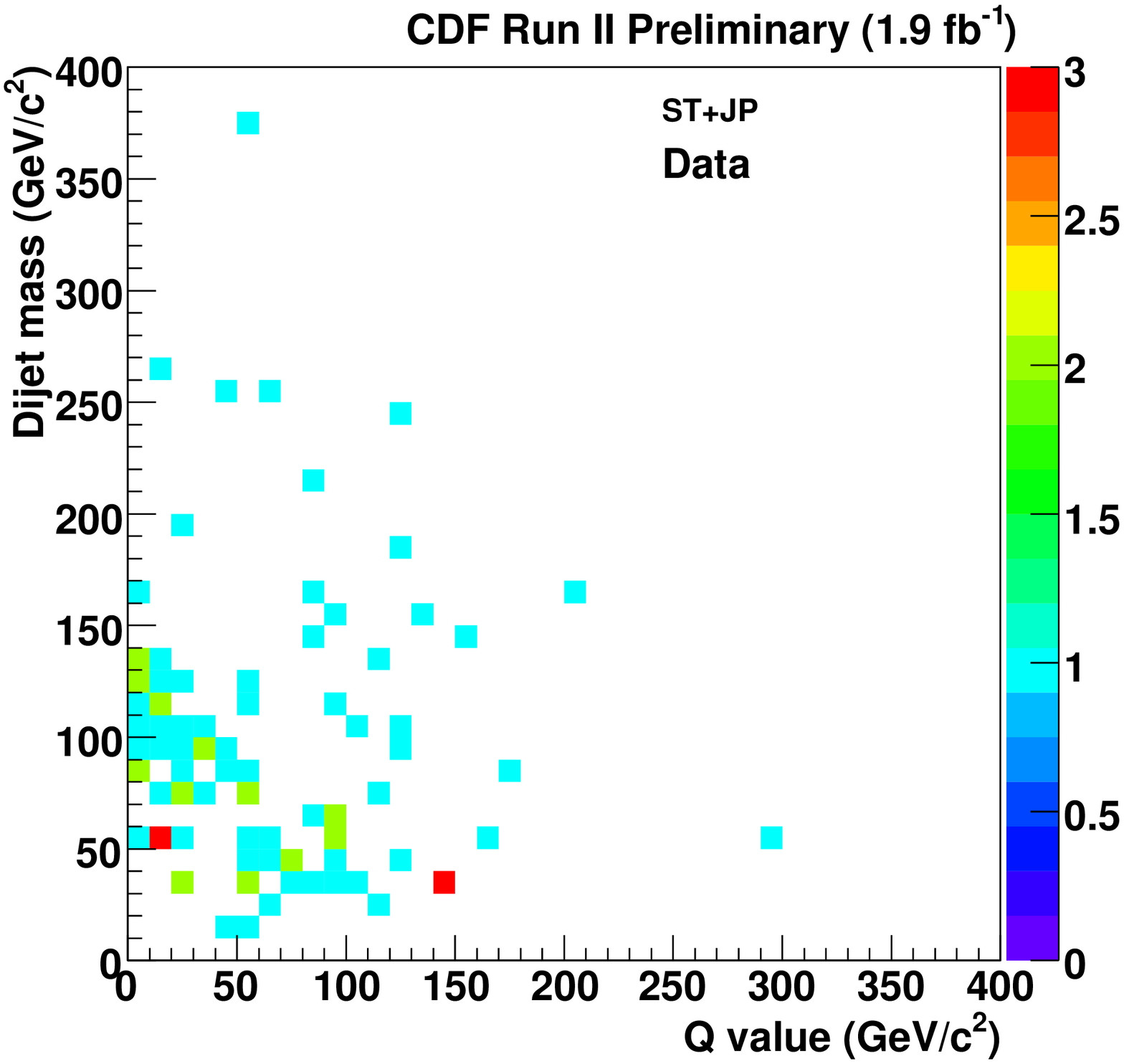}
    \includegraphics[width=0.25\textwidth]{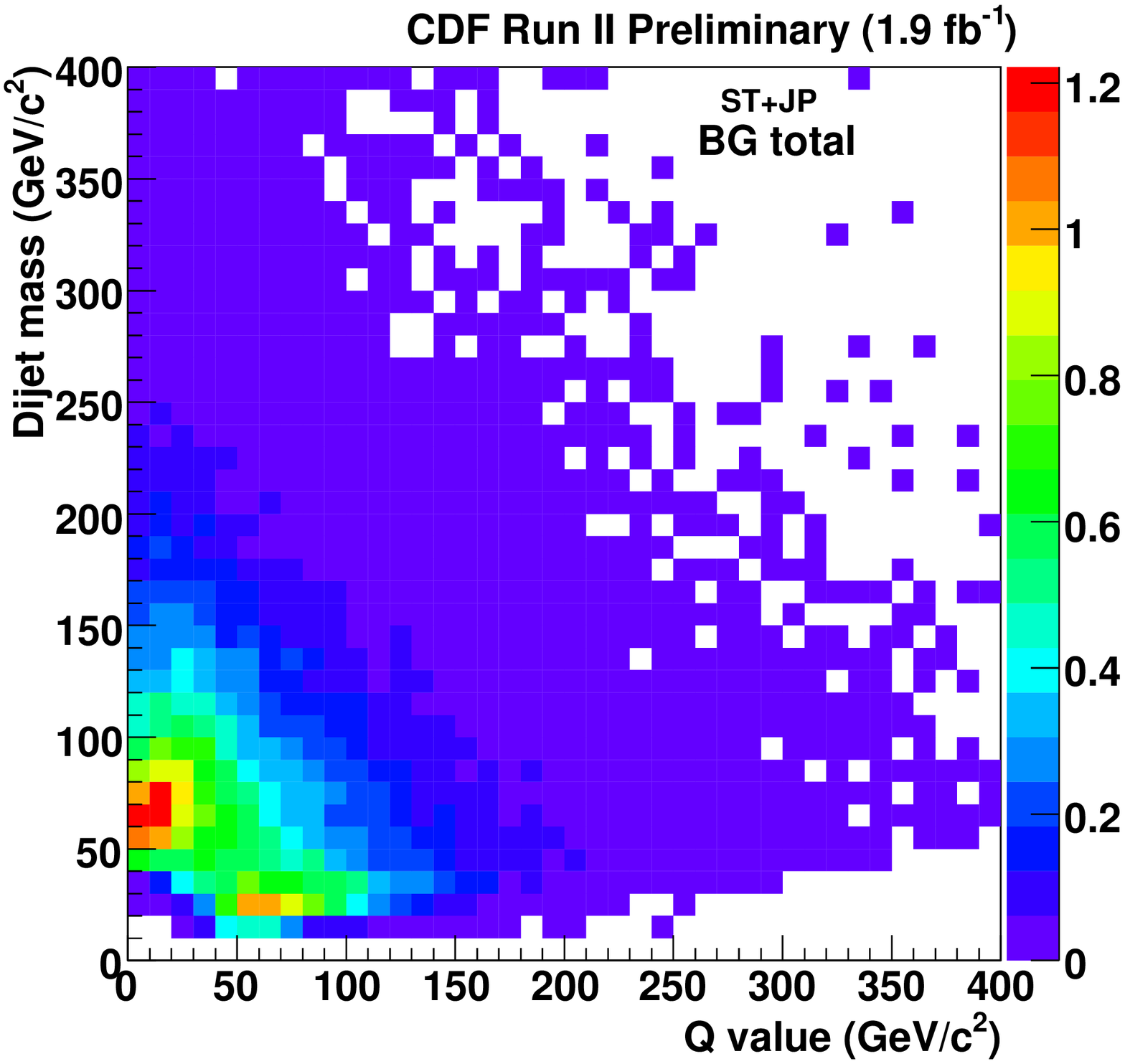}
    \includegraphics[width=0.25\textwidth]{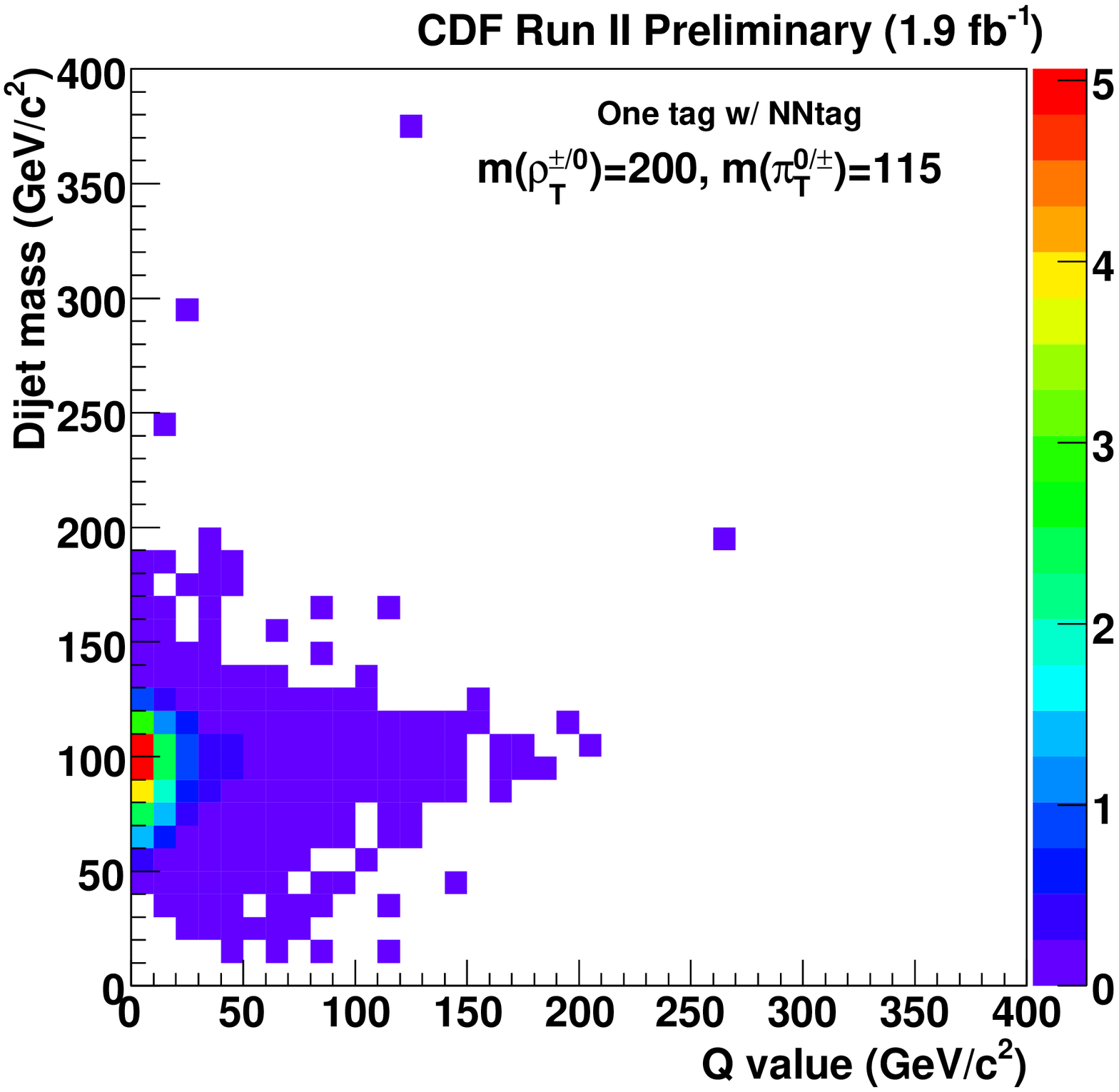}
    \includegraphics[width=0.25\textwidth]{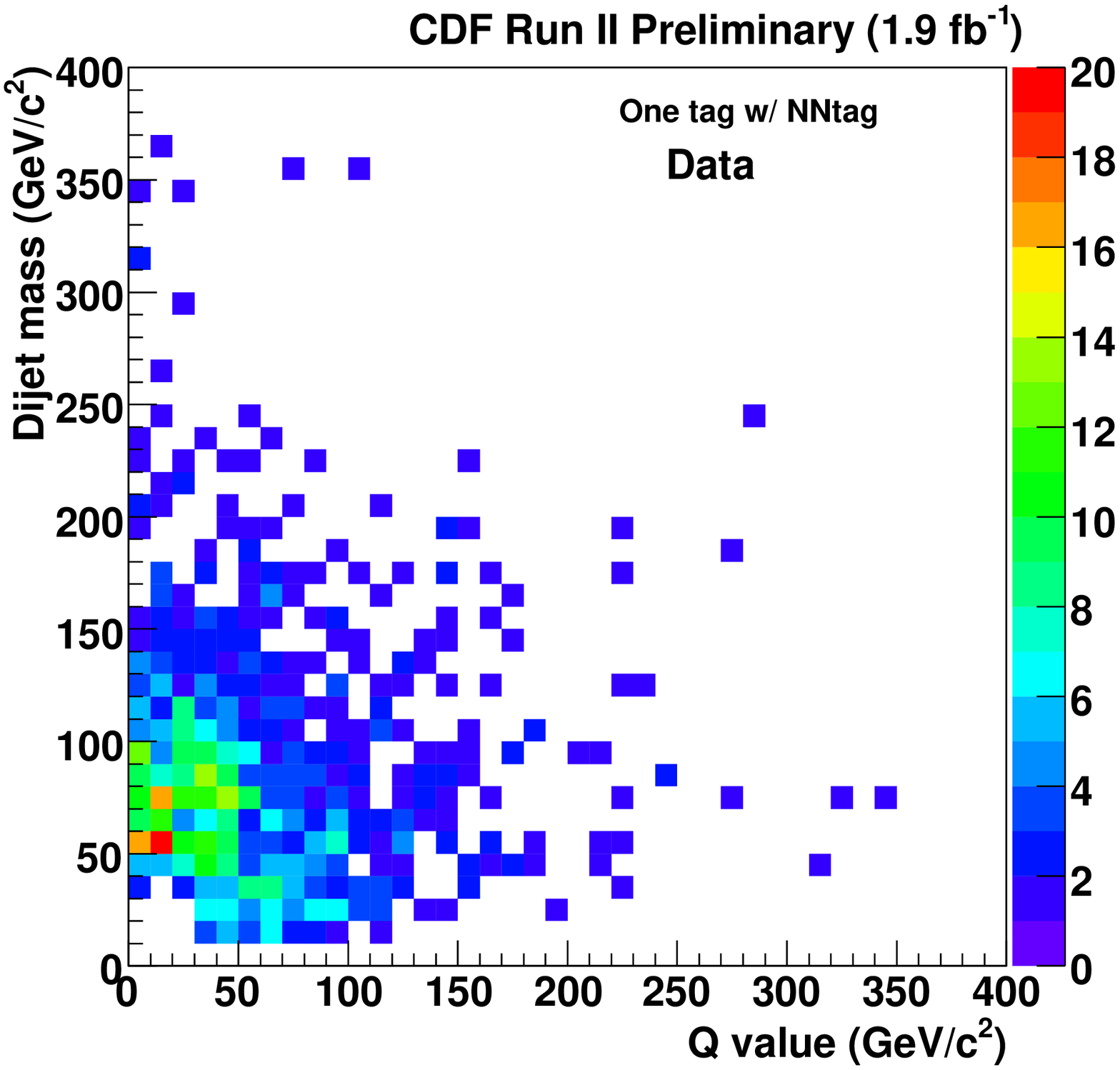}
    \includegraphics[width=0.25\textwidth]{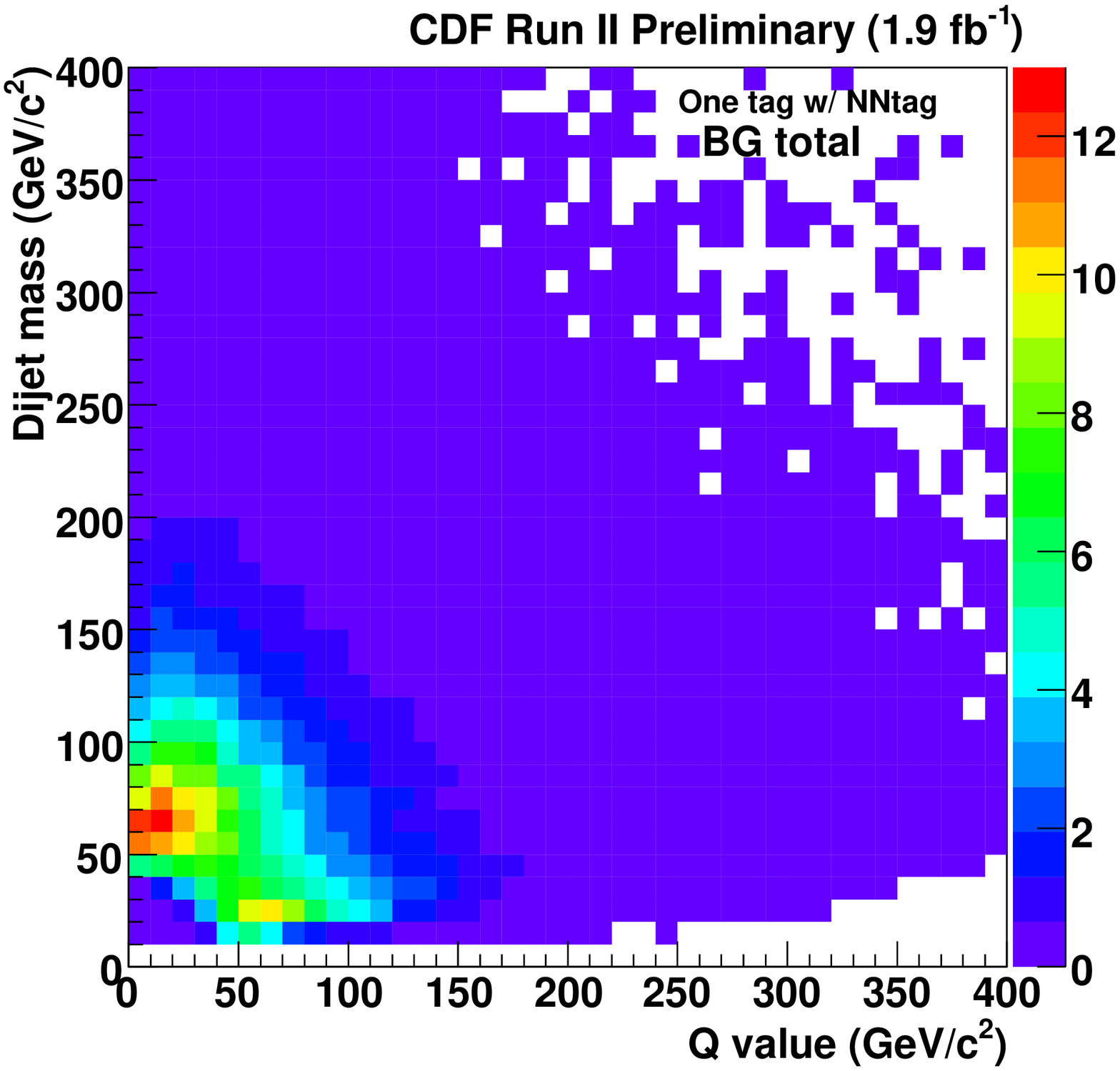}
    \caption{Dijet mass vs Q value 2-dimensional distribution of 
      signal MC (left column, $m(\rho_T) = 200 \mathrm{GeV/c^2}$, $m(\pi_T) = 115 \mathrm{GeV/c^2}$), 
      data (middle column) and backgrounds (right column) for the each tagging categories.
      Top line three plots are the double secondary vertex tag category, 
      middle line are the one secondary vertex tag + one jet probability tag category and 
      bottom line are the one secondary vertex tag with NN tag category, respectively.}
     \label{fig:Mjj_vs_Qval}
  \end{center}
\end{figure}

Since there is no significant excess of events in the data compared to the predicted background,
we set the 95\% C.L. excluded region on technicolor production 
as a function of the technicolor particles mass.
A 2-dimensional binned maximum likelihood technique is used on the 2-dimensional distribution 
of dijet invariant mass vs $Q$-value by constraining the number of background events 
within the uncertainties, where $Q$-value is defined as $Q = m(\rho_T) - m(\pi_T) - m(W)$.
Figure.~\ref{fig:Mjj_vs_Qval} shows 
the 2-dimensional distribution of data, backgrounds and signal for each tag category.

\begin{figure}[htbp]
  \begin{center}
    \includegraphics[width=0.45\textwidth]{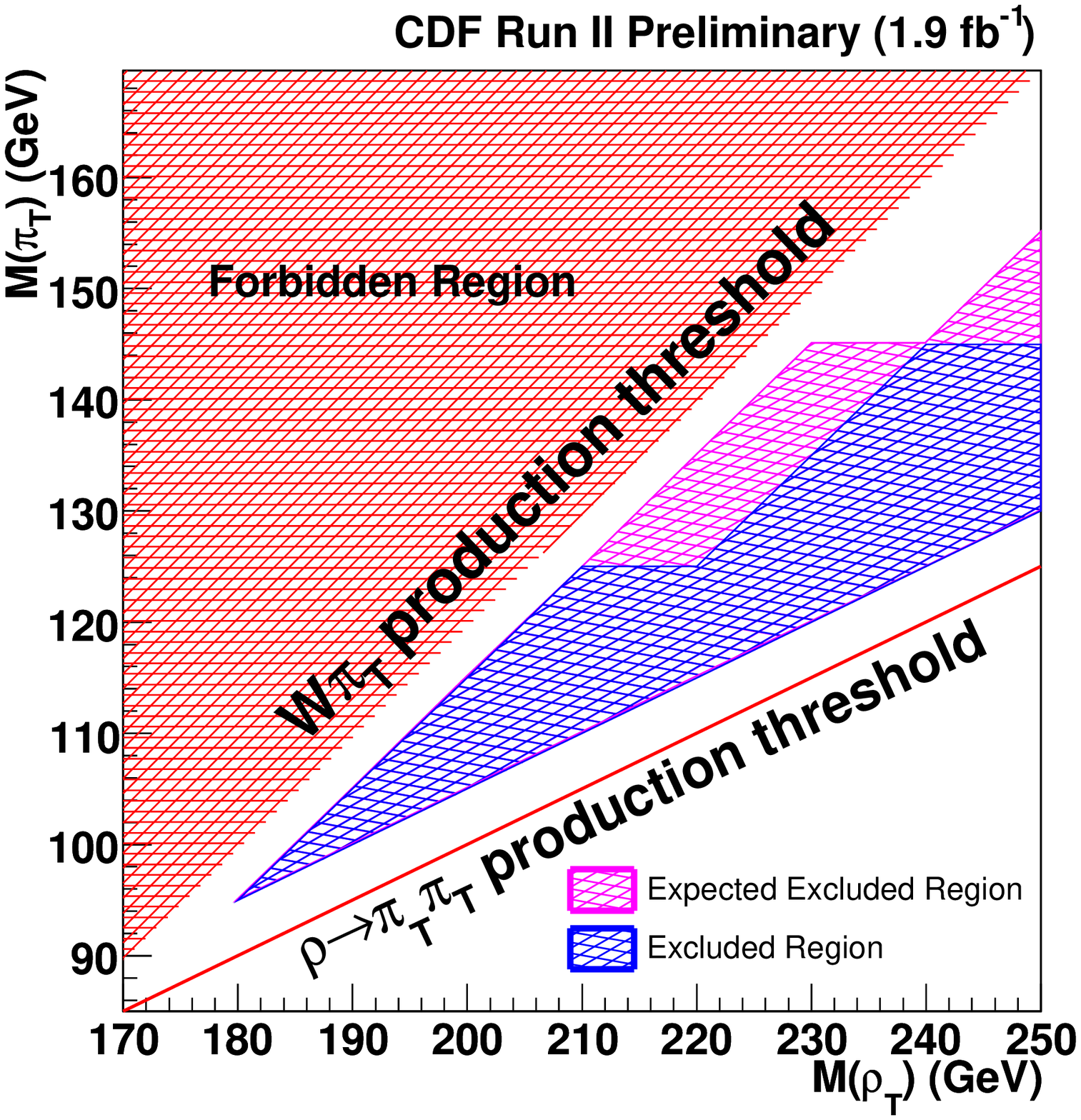}
    \caption{The expected and observed excluded region at 95\% C.L. as a function of technirho mass and technipi mass.
Kinematical threshold of $W \pi_T$ and $\rho_T \rightarrow \pi_T \pi_T$ are shown in the figure.
}
    \label{fig:Exclude}
  \end{center}
\end{figure}

The final expected and observed excluded region at 95\% C.L. are shown in Figure~\ref{fig:Exclude}. 
A region of $m(\rho_T) = 180$ - $250 \mathrm{GeV}$ are excluded at 95\% C.L. 
based on the Technicolor Straw Man Model, 
except the region which are near the $W \pi_T$ production threshold 
with $m(\rho_T) \geq 220 \mathrm{GeV}$ and $m(\pi_T) \geq 125 \mathrm{GeV}$.

\begin{acknowledgments}
We thank Ken Lane for helpful discussions, the Fermilab staff and the technical staffs of the
participating institutions for their vital contributions. This
work was supported by the U.S. Department of Energy and National
Science Foundation; the Italian Istituto Nazionale di Fisica
Nucleare; the Ministry of Education, Culture, Sports, Science and
Technology of Japan; the Natural Sciences and Engineering Research
Council of Canada; the National Science Council of the Republic of
China; the Swiss National Science Foundation; the A.P. Sloan
Foundation; the Bundesministerium f\"ur Bildung und Forschung,
Germany; the Korean Science and Engineering Foundation and the
Korean Research Foundation; the Particle Physics and Astronomy
Research Council and the Royal Society, UK; the Russian Foundation
for Basic Research; the Comision Interministerial de Ciencia y
Tecnologia, Spain; and in part by the European Community's Human
Potential Programme under contract HPRN-CT-20002, Probe for New
Physics.
\end{acknowledgments}

\end{document}